\documentclass[aps,prd,reprint,longbibliography,superscriptaddress,nofootinbib,floatfix]{revtex4-1}
\pdfoutput=1

\usepackage{amsfonts,amssymb,amsmath}
\usepackage{epsfig}
\usepackage{hyperref}
\hypersetup{colorlinks   = true,
            urlcolor     = blue,
            citecolor    = blue,
            linkcolor    = blue,
            menucolor    = blue,
            anchorcolor  = blue,
            filecolor    = blue
}

\enlargethispage{\baselineskip}

\smallskipamount

\everymath{\displaystyle}

\begin{document}

\title{Gravitational Signatures of Axion Dark Matter via Parity-Violating Interactions}

\newcommand{\Unisa}{\affiliation{Dipartimento di Fisica ``E.R.\ Caianiello'', Universit\`a degli Studi di Salerno,\\ Via Giovanni Paolo II, 132 - 84084 Fisciano (SA), Italy}}
\newcommand{\INFN}{\affiliation{Istituto Nazionale di Fisica Nucleare - Gruppo Collegato di Salerno - Sezione di Napoli,\\ Via Giovanni Paolo II, 132 - 84084 Fisciano (SA), Italy}}

\author{Marco Figliolia}
\email{mfigliolia@unisa.it}
\Unisa \INFN
\thanks{MF and FG contributed equally to this work}

\author{Francesco Grippa}
\email{fgrippa@unisa.it}
\Unisa \INFN
\thanks{MF and FG contributed equally to this work}

\author{Gaetano Lambiase}
\email{glambiase@unisa.it}
\Unisa \INFN

\author{Luca Visinelli}
\email{lvisinelli@unisa.it}
\Unisa \INFN

\begin{abstract}
We investigate axion-like particles coupled to gravity through a parity-violating Chern--Simons (CS) interaction. In this framework, axion dark matter (DM) can decay into pairs of circularly polarized gravitons, producing a persistent, nearly monochromatic GW signal. We compute the expected signal at Earth assuming a Navarro–Frenk–White Galactic halo model with the corresponding velocity distribution, and compare it with the narrowband sensitivities of the LIGO O4 run and the projected reach of the Einstein Telescope. The resulting bounds on the axion–graviton coupling $\alpha$ improve upon the cosmological stability requirement for axion masses $m_\phi \lesssim 10^{-11}$\,eV, excluding values up to four orders of magnitude below the stability limit. This constitutes a robust direct terrestrial constraint on the axion–gravity CS coupling. We also discuss distinctive observational signatures, such as circular polarization asymmetries, annual modulation, and potential enhancements from DM substructures, which could serve as smoking-gun evidence for parity-violating gravitational interactions.
\end{abstract}

\maketitle

\section{Introduction}
\label{sec:introduction}

Dark matter (DM) is one of the clearest indications for physics beyond the Standard Model (SM), yet its fundamental nature remains unknown. Among the leading candidates are axions and axion-like particles, pseudoscalar fields that arise generically from the spontaneous breaking of global symmetries~\cite{Weinberg:1972fn, Peccei:1977hh, Weinberg:1977ma, Wilczek:1977pj} and in string theory compactifications~\cite{Witten:1984dg, Svrcek:2006yi, Choi:2006za, Arvanitaki:2009fg, Cicoli:2012sz, Gendler:2023kjt}. Their feeble couplings to visible matter, together with the natural production of a cosmic condensate through the vacuum misalignment mechanism, make axions viable DM candidates over a broad mass range~\cite{Preskill:1982cy, Abbott:1982af, Dine:1982ah}. In this framework, the axion field behaves as a coherently oscillating background that redshifts as cold DM once the Hubble rate drops below the axion mass~\cite{Turner:1985si, Sikivie:2006ni, Marsh:2015xka}. This mechanism can naturally yield the observed DM abundance for a wide range of decay constants and initial misalignment angles, and axions can in fact account for the entirety of the DM density~\cite{Hertzberg:2008wr, Visinelli:2009zm, Arias:2012az}.

Most phenomenological work has concentrated on axion couplings to the electromagnetic (EM) sector, which is central to laboratory searches based on axion--photon conversion~\cite{Sikivie:1983ip, Sikivie:1985yu}, affects stellar energy-loss arguments~\cite{Raffelt:1987yu}, and can produce cosmic birefringence signatures in the CMB and polarised sources~\cite{Choi:2021aze, Lin:2025gne, Ballardini:2025apf}. A broad program of laboratory experiments, astrophysical observations, and cosmology has placed stringent limits on EM couplings as well as SM fermions such as electrons and nucleons~\cite{Graham:2015ouw, DiLuzio:2020wdo, Irastorza:2018dyq}. Recent targeted radio searches for transient signatures demonstrate the complementary reach of astrophysical probes in parameter regions inaccessible to laboratory experiments~\cite{Walters:2024vaw}. In models where couplings to the SM are suppressed (e.g. photophobic or hidden-sector constructions), gravitational channels can become the leading observable channel, motivating a systematic study of their signatures~\cite{Craig:2018kne, Bauer:2021mvw}. A particularly well-motivated example is the dimension-five, parity-odd Chern--Simons (CS) operator
\begin{equation}
    \label{eq:r_rtilde}
    \mathcal{L}_{\mathrm{CS}} \supset \alpha \phi R \widetilde{R}\,,\qquad R \widetilde{R} \equiv \frac{1}{2} \epsilon^{\mu\nu\alpha\beta} R_{\mu\nu\rho\sigma} {R_{\alpha\beta}}^{\rho\sigma}\,,
\end{equation}
where $R_{\mu\nu\rho\sigma}$ is the Riemann tensor, $\epsilon^{\mu\nu\alpha\beta}$ the Levi-Civita tensor, and $\alpha$ is the coupling constant with dimensions of length in natural units. The CS operator appears in a variety of theoretical contexts, including anomaly-cancellation and Green--Schwarz-type couplings in string constructions~\cite{Green:1984sg}, higher-curvature corrections in string-inspired models~\cite{Zwiebach:1985uq}, and more general extensions of general relativity (GR)~\cite{Jackiw:2003pm, Alexander:2009tp}. For a strictly homogeneous and static scalar background, the CS term reduces to a total derivative and is dynamically irrelevant. On the other hand, a spacetime-dependent background field, such as a time-varying DM condensate, breaks discrete parity symmetry. This induces parity--violating effects in gravitational wave (GW) propagation, such as birefringence~\cite{Alexander:2004wk}, helicity-dependent dispersion relations~\cite{Daniel:2024lev}, and radiative signatures in curved spacetime. Early work using results by the Laser Interferometer Gravitational-wave Observatory (LIGO) established constraints on the CS coupling via resonant GW enhancement~\cite{Jung:2020aem}, gravitational birefringence and modified propagation speeds~\cite{Chu:2020iil}, effects on neutrino propagation~\cite{Lambiase:2022ucu}, and resonant amplification from axion clouds in CS gravity~\cite{Fujita:2020iyx}. In the early universe, axion CS couplings can source stochastic GW backgrounds through parametric resonance, potentially detectable across a wide range of frequency bands~\cite{Li:2023vuu}. Such couplings may also amplify the stochastic background~\cite{Alexander:2024klf}.

Direct bounds on gravitational couplings of the form in Eq.~\eqref{eq:r_rtilde} remain comparatively less constrained by terrestrial experiments because of their Planck-suppressed strength and the historical absence of sensitive probes. The advent of GW astronomy changes this picture: current interferometry searches at LIGO~\cite{KAGRA:2025yfg} have already begun to test parity-violating propagation effects~\cite{Yunes:2010yf, Alexander:2017jmt, Yamada:2020zvt, Okounkova:2021xjv}, and future facilities like the Einstein Telescope (ET)~\cite{ET:2019dnz,Abac:2025saz} will greatly improve sensitivity, particularly at low frequencies relevant for ultra-light fields.

In this paper we study the scenario where the CS interaction dominates the low-energy coupling between an axion-like DM field and gravity. We compute the GW power directly emitted by axion decay into two gravitons via the CS operator, estimate the resulting Galactic graviton flux at Earth, and confront the predicted signal with current and projected interferometer sensitivities to derive bounds on the coupling strength. We find that for ultra-light masses $m_{\phi}\lesssim 10^{-11}\,$eV the direct GW limits improve upon cosmological stability bounds in relevant regions of parameter space. To our knowledge, these results provide among the first direct constraints on axion--graviton CS couplings derived from terrestrial GW data, complementing previous studies that focused on birefringence and early-universe production~\cite{Hook:2016mqo, Craig:2018kne}. The remainder of this work is organized as follows. In Sec.~\ref{sec:methods}, we review the theoretical framework and the mechanism for axion decay into gravitons, establishing the conditions under which this channel dominates. Section~\ref{sec:results} presents the resulting constraints on the axion–graviton coupling, derived from the Galactic graviton flux and compared against both cosmological stability bounds and interferometer sensitivities. A discussion of the implications is given in Sec.~\ref{sec:discussion}, while results are summarized in Sec.~\ref{sec:conclusions}. Throughout, we adopt natural units with $c=\hbar=1$ and $G = 1/m_{\rm Pl}^2$.

\section{Methods}
\label{sec:methods}

\subsection{Theoretical Framework}

In this section, we outline the theoretical framework adopted in our analysis. The axion field $\phi$ is treated as a fixed external background, while gravitational perturbations are quantized around a classical spacetime. The total Lagrangian describing the axion and its interactions with the SM and gravity is given by~\cite{Alexander:2009tp}
\begin{equation}
    \begin{split}
    \mathcal{L} &= \mathcal{L}_\mathrm{EH} + \mathcal{L}_\phi + \mathcal{L}_\mathrm{CS} + \mathcal{L}_{\phi\gamma\gamma} + \mathcal{L}_{\phi\psi\psi} \\
    &= \frac{1}{16 \pi G} R - \left[ \frac{1}{2} \nabla^\mu \phi \nabla_\mu \phi - V(\phi) \right] + \alpha \, \phi R \tilde{R}\\ 
    &+ \frac{1}{4 f_\phi} \, \phi F_{\mu\nu} \widetilde{F}^{\mu\nu} + i g_\psi \, \phi \, \bar{\psi} \gamma^5 \psi \,,
    \end{split}
\end{equation}
where $R$ is the Ricci scalar, $R \tilde{R}$ the Pontryagin density, $F_{\mu\nu}$ the EM field strength, and $\psi$ denotes SM fermions. The parameters $f_\phi$ and $g_\psi$ quantify the axion-photon and axion-fermion couplings, respectively. Since $R \tilde{R}$ is parity-odd, $\phi$ must be a pseudo-scalar to preserve parity invariance. The axion--photon and axion--fermion interactions induce the decay rates~\cite{DiLuzio:2020wdo}
\begin{equation}
    \Gamma_{\phi \rightarrow \gamma\gamma} = \frac{g_\gamma^2\,m_\phi^3}{64 \pi f_\phi^2}\,, \qquad \Gamma_{\phi \rightarrow \psi\psi} = \frac{g_\psi^2 m_\phi}{8 \pi} \left( 1 - \frac{4 m_\psi^2}{m_\phi^2} \right)^{1/2}\,,
\end{equation}
where $g_\gamma^2$ is a dimensionless constant that quantifies the strength of the axion coupling to photons. For the purpose of computing decay rates, we treat fluctuations of $\phi$ as a particle degree of freedom, though in the DM context its background oscillations can be approximated as a classical condensate.

We first discuss the axion-fermion interaction. For an ultra-light axion, decays into massive SM fermions are kinematically forbidden, except possibly for massless neutrinos of mass $m_\nu \ll m_\phi$. Even in that case, the neutrino Yukawa coupling scales as $g_\nu \propto m_\nu / f_\phi$, making the decay highly suppressed. Consequently, the photon decay channel is often dominant in generic QCD axion models~\cite{Bauer:2017ris}, although additional channels exist, including fermion pairs~\cite{Bauer:2021mvw, DiLuzio:2024jip}, gluons~\cite{Chakraborty:2021wda, Bisht:2024hbs}, and electroweak bosons~\cite{Alonso-Alvarez:2018irt, Aiko:2023trb}. For light axions with $m_\phi \ll m_W, m_Z$, all of these channels are kinematically suppressed. Nevertheless, this hierarchy is not universal. In photophobic axion models featuring an additional left-right (LR) chiral symmetry $SU(2)_L \times SU(2)_R$, the axion does not couple to SM photons at tree level. Furthermore, being CP-odd and LR-even, axion-fermion couplings are forbidden unless these symmetries are broken~\cite{Brivio:2017ije}. For axion masses in the range $100\,\mathrm{GeV} \le m_\phi \le 4000\,\mathrm{GeV}$, the decay pattern is dominated by $\phi \rightarrow W^+W^-$ ($\sim 70\%$), with subleading channels $\phi \rightarrow \gamma Z$ ($\sim 20\%$) and $\phi \rightarrow ZZ$ ($\sim 10\%$)~\cite{Ding:2024djo}.

Scenarios with a suppressed axion--photon coupling $g_\gamma \ll 1$, arising for instance from accidental cancellations in the EM anomaly coefficient or if the axion originates in a hidden sector with no direct coupling to the visible photon, provide a well-motivated context in which the gravitational CS decay in Eq.~\eqref{eq:r_rtilde} can dominate. The corresponding decay into gravitons is~\cite{Delbourgo:2000nq}
\begin{equation}
    \label{eq:phiggrate}
    \Gamma_{\phi \rightarrow gg} = \frac{\pi \, G^2 m_\phi^7}{2} \, \alpha^2\,.
\end{equation}
In these cases, the axion effectively behaves as a feebly interacting particle, with GWs as its dominant decay products~\cite{Hook:2016mqo, Craig:2018kne}. Note, that the exponent in the ultralight axion mass in Eq.~\eqref{eq:phiggrate} makes the DM condensate cosmologically stable and extraordinarily long-lived. The axion lifetime is then given by the competition between the $\phi \rightarrow gg$ and $\phi\rightarrow\gamma\gamma$ channels,
\begin{equation}
    \tau_\phi = \min\!\left( \frac{1}{\Gamma_{\phi \rightarrow \gamma\gamma}}, \frac{1}{\Gamma_{\phi \rightarrow gg}} \right)\,.
\end{equation}
In this work, we focus on scenarios in which $\Gamma_{\phi \rightarrow gg} \gtrsim \Gamma_{\phi \rightarrow \gamma\gamma}$, allowing a clean extraction of constraints on the axion-graviton coupling without introducing additional parameters. This requires the parameter constraint
\begin{equation}
    \alpha f_\phi\gtrsim \frac{g_\gamma}{4\sqrt{2}\pi}\frac{m_{\rm Pl}^2}{m_\phi^2} \sim 10^{77}\,g_\gamma\,\left(\frac{10^{-11}{\rm\,eV}}{m_\phi}\right)^2\,.
\end{equation}
This inequality ensures that the gravitational decay channel dominates over the photon channel. Alternatively, the CS interaction is parameterised by introducing a fundamental length scale $\ell_{\rm CS}$, such that the effective coupling reads $\alpha = \ell_{\rm CS}^2/(8\sqrt{8\pi G})$~\cite{Lambiase:2022ucu}. The condition that the gravitational decay $\phi \to gg$ competes with the EM decay $\phi \to \gamma\gamma$ is then
\begin{equation}
    \label{eq:requirement}
    m_\phi \,\ell_{\rm CS} \gtrsim \frac{2}{\pi^{1/4}} \left( \frac{g_\gamma\,m_{\rm Pl}}{f_\phi} \right)^{1/2}\,.
\end{equation}
This requirement is controlled by the dimensionless combination $m_\phi \ell_{\rm CS}$, with only a mild dependence on the axion decay constant $f_\phi$. For $f_\phi = 10^{12}$\,GeV, one finds $m_\phi \ell_{\rm CS} \gtrsim 5\times 10^3\,g_\gamma^{1/2}$, so that for an axion mass $m_\phi = 10^{-12}$\,eV the bound gives $\ell_{\rm CS} \gtrsim 10^9$\,m, of order the solar radius. At the same time, causality and perturbative unitarity of the low-energy effective field theory (EFT) impose an upper bound on the dimensionless combination $m_\phi \,\ell_{\rm CS} \lesssim 1$~\cite{Serra:2022pzl}, which, once combined with the requirement in Eq.~\eqref{eq:requirement} gives an upper bound on the axion–photon coupling,
\begin{equation}
    g_\gamma \lesssim \mathcal{O}\!\left(\frac{f_\phi}{m_{\rm Pl}}\right)\,.
\end{equation}
Therefore, for the gravitational decay $\phi\to gg$ to dominate while remaining inside the EFT window one naturally requires a strongly suppressed axion–photon coupling. In the photophobic limit this condition is easily satisfied within the EFT; non-photophobic realizations would instead demand an explicit UV completion that generates the effective Chern–Simons coupling while maintaining causality and unitarity at higher energies.

While the unitarity condition $m_\phi \ell_{\rm CS} \lesssim 1$ marks the regime in which the low-energy EFT remains self-consistent, this limit does not necessarily affect the underlying theory. Our calculation applies to any framework in which an operator of the form in Eq.~\eqref{eq:r_rtilde} arises with an effective coupling $\alpha$, regardless of whether the low-energy expansion remains valid up to this bound. Large effective couplings may originate from UV completions that reinstate higher-order operators or from non-perturbative dynamics beyond the EFT description. Therefore, the constraints derived in this work should be interpreted as phenomenological bounds on the effective CS interaction, similarly to astrophysical limits on the axion-photon coupling.

Known UV completions generating the CS operator predict very small effective couplings, $\alpha \simeq 10^{-21}\!-\!10^{-16}\,\mathrm{GeV}^{-1}$, as found in string-inspired constructions~\cite{Cicoli:2012sz,Svrcek:2006yi}, quantum gravity scenarios~\cite{Taveras:2008yf}, and fermionic completions~\cite{Alexander:2022cow}. In string-motivated models, $\alpha$ is usually suppressed by either the Planck scale or the axion decay constant, leading to $\alpha \sim 10^{-18\text{--}16}\,\mathrm{GeV}^{-1}$, while in quantum gravity approaches one obtains $\alpha \simeq M_{\rm Pl}^{-1} \approx 10^{-19}\,\mathrm{GeV}^{-1}$. Although these values lie far below the region probed in our analysis, our constraints remain phenomenologically relevant, as they apply to any theory that yields the same effective CS operator, independently of its microscopic origin.

Although large values of the CS coupling may appear exotic from an effective field theory perspective, they illustrate that the required enhancement of the gravitational channel can be expressed in terms of a moderate dimensionless condition rather than an extreme fine-tuning. Photophobic or hidden-sector models, where EM couplings are naturally suppressed, might naturally accommodate phenomenologically viable scenarios where CS-mediated decays dominate. In the next section we compute the resulting Galactic graviton flux from axion decay, and compare the signal strength to the sensitivities of current and future interferometers.

\subsection{Graviton flux from axion decay}
\label{sec:graviton_flux}

Axions gravitationally bound in the Milky Way halo can decay into pairs of gravitons, producing a monochromatic GW signal with frequency set by the axion mass. The emitted gravitons propagate through the Galaxy and reach Earth as an effectively isotropic and stationary background. The differential graviton flux at Earth from axions of mass $m_\phi$ and lifetime $\tau_\phi$ is given by~\cite{Arguelles:2022nbl}\footnote{We omit the factor $1/3$ that accounts for neutrino flavor mixing.}
\begin{equation}
    \frac{{\rm d} \Phi_g}{{\rm d}E} = \frac{1}{4 \pi} \frac{1}{\tau_\phi m_\phi} \frac{{\rm d}N_g}{{\rm d}E} D(\Delta\Omega)\,,
\end{equation}
where ${\rm d}N_g/{\rm d}E$ is the energy spectrum of gravitons in the decay, and $D(\Delta\Omega)$ is the line-of-sight integral of the axion density $\rho_\phi(r)$ over the solid angle $\Delta\Omega$:
\begin{equation}
    D(\Delta\Omega) = \int_{\Delta\Omega} {\rm d}\Omega \int_{\rm l.o.s.} \rho_\phi (r)\,{\rm d}\ell\,.
\end{equation}
For the two-body decay $\phi \rightarrow gg$, the rest-frame spectrum is then
\begin{equation}
    \frac{{\rm d}N_g}{{\rm d}E} = 2\,\delta\!\left(\frac{m_\phi}{2} - E\right)\,.
\end{equation}
We model the Galactic axion distribution using a Navarro-Frenk-White (NFW) profile~\cite{Navarro:1996gj},
\begin{equation}
    \rho_\phi(r) = \rho_0 \frac{r_s/r}{(1+r_s/r)^2}\,,
\end{equation}
where $r_s = 20$\,kpc is the scale radius, and $\rho_0$ is fixed by the solar DM density $\rho_\odot = 0.4{\rm\,GeV\,cm^{-3}}$~\cite{Salucci:2010qr, deSalas:2019pee}. The solar distance to the Galactic Center is $r_\odot \simeq 8.5{\rm\,kpc}$. The line-of-sight coordinate $\ell$ and the Galactocentric radius $r$ are related by
\begin{equation}
    \label{eq:l_to_r}
    \ell^2(r) = r_\odot^2 + r^2 - 2 \, r_\odot \, r \cos\psi \,,
\end{equation}
where $\psi$ is the angle between the pointing direction and the Galactic Center. This allows the $D$-factor to be evaluated numerically in terms of $r$.

To relate the graviton flux to a observable strain, we use linearized GR. In transverse-traceless gauge, the GW energy flux is
\begin{equation}
    p_\mathrm{GW} \equiv \left\langle \frac{{\rm d}E_\mathrm{GW}}{{\rm d}A\, {\rm d}t} \right\rangle = \frac{\pi}{4 G} \,f^2 h^2 \,,
\end{equation}
where $h$ is the strain amplitude and $f = m_\phi/(4 \pi)$ is the GW frequency corresponding to the axion mass. Equating the GW energy flux to the integrated graviton flux,
\begin{equation}
    \int E \, \frac{{\rm d}\Phi_g}{{\rm d}E} \, {\rm d}E = p_\mathrm{GW}\,,
\end{equation}
yields the strain induced by axion decay through the CS operator:
\begin{equation}
    \label{eq:strain}
    h = \left(\frac{D(\Delta\Omega) \alpha^2 G^3 m_\phi^7}{2 f^2}\right)^{1/2}\,.
\end{equation}
This expression allows a direct translation from interferometer strain limits into constraints on the axion--graviton coupling $\alpha$.

\section{Results}
\label{sec:results}

For a given axion mass, the strain predicted by Eq.~\eqref{eq:strain} can be directly compared with the LIGO O4 limits~\cite{KAGRA:2025yfg} and the projected sensitivities of ET~\cite{Hild:2010id, Abac:2025saz}. This yields an upper bound on the axion--graviton coupling,
\begin{equation}
    \label{eq:boundalpha}
    \alpha \lesssim \left(\frac{2h^2}{G^3 m_\phi^7 D(\Delta\Omega)}\right)^{1/2}\,.
\end{equation}
valid in regimes where axions decay predominantly into gravitons, such as photophobic models.

LIGO reports strain sensitivity as an amplitude spectral density, $S_h^{1/2}(f)$, in units of $1/\sqrt{\rm Hz}$. To compare the GW reach with the nearly monochromatic axion signal, one must fold in the intrinsic linewidth set by the Galactic velocity dispersion,
\begin{equation}
    \frac{\Delta f_{\rm gal}}{f} \sim 10^{-3} \,.
\end{equation}
The corresponding effective strain sensitivity is
\begin{equation}
    \label{eq:reach}
    h \approx S_h^{1/2}(f) \; \sqrt{\Delta f_\mathrm{signal}} \,,
\end{equation}
where $\Delta f_\mathrm{signal} = \Delta f_{\rm gal}$. Equation~\eqref{eq:reach} translates the detector sensitivity into a strain limit for the axion-induced signal at Earth. Substituting this expression into Eq.~\eqref{eq:boundalpha} leads to the maximum allowed $\alpha$ consistent with the absence of a spectral line at $f = m_\phi/(4\pi)$. Axions that constitute a non-negligible fraction of the present-day DM density must be cosmologically stable. This requirement imposes the cosmological \emph{stability bound}, obtained by demanding that the axion lifetime exceeds the age of the Universe,
\begin{equation}
    \tau_\phi > t_\mathrm{univ} \simeq 13.8 \times 10^9{\rm\,years} \,,
\end{equation}
which translates into a bound valid for photophobic axions as\footnote{In non-photophobic scenarios, the effective bound is weakened according to the branching ratio into gravitons, ${\rm Br}(\phi\to gg)$, so that Eq.~\eqref{eq:stability} results in $\alpha \to \alpha/ \sqrt{{\rm Br}(\phi\to gg)}$. Practically, the bound is placed over the quantity $\alpha/ \sqrt{{\rm Br}(\phi\to gg)}$.}
\begin{equation}
    \label{eq:stability}
    \alpha < \left(\frac{2}{\pi}\frac{1}{G^2 \, t_\mathrm{univ} \, m_\phi^7}\right)^{1/2}\,.
\end{equation}

\begin{figure}[htb]
    \centering
    \includegraphics[width=\linewidth]{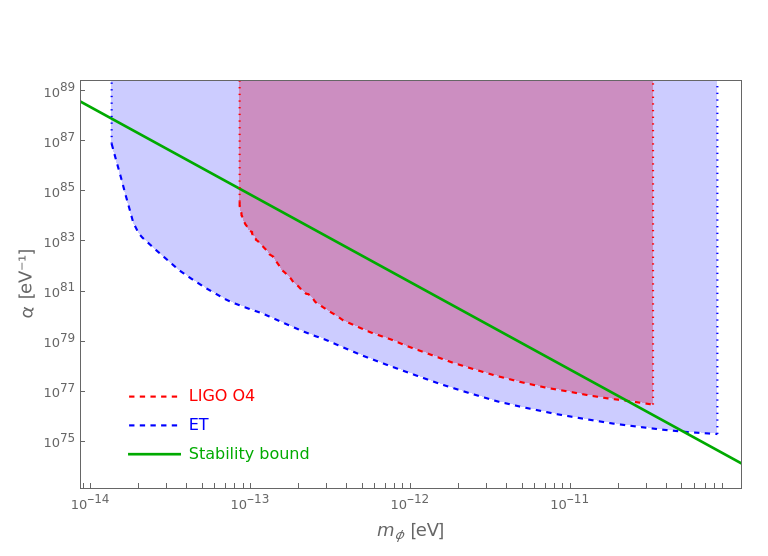}
    \caption{Axion-graviton coupling $\alpha$ as function of the axion mass $m_\phi$. The green solid line denotes the cosmological stability bound in Eq.~\eqref{eq:stability}. The red shaded region is excluded by the non-detection of a monochromatic signal in LIGO O4 data, and the blue shaded region shows the projected exclusion reach of ET. Ground-based interferometers improve upon the stability bound by several orders of magnitude for $10^{-14} \lesssim m_\phi/{\rm eV} \lesssim 10^{-11}$.}
    \label{fig:bounds_interpolation}
\end{figure}

Figure~\ref{fig:bounds_interpolation} shows the resulting constraints. The green curve denotes the cosmological stability limit, while the red and blue shaded regions are excluded by LIGO O4 and the projected ET sensitivity, respectively. At large masses, $m_\phi \gtrsim 10^{-11}$\,eV, the stability condition dominates. In contrast, at smaller masses, $10^{-14} \lesssim m_\phi/{\rm eV} \lesssim 10^{-11}$, interferometers improve upon the stability requirement by several orders of magnitude. For example, at $m_\phi \sim 10^{-12}$\,eV the LIGO O4 data exclude couplings as strong as $\alpha \lesssim \mathcal{O}(10^{78}{\rm\,eV^{-1}})$, approximately four orders of magnitude below the stability limit. The ET sensitivity extends this reach even further, with projected bounds tightening by up to three additional orders of magnitude in the same mass window. This scaling follows from Eq.~\eqref{eq:boundalpha}, where the strain sensitivity directly translates into $\alpha \propto h/m_\phi^{7/2}$.

These results highlight two important features. First, ground-based interferometers are already sensitive to axion–graviton couplings at the level of cosmological stability, thereby establishing the first direct terrestrial constraints on the CS interaction. Second, the predicted signal is nearly monochromatic, with a linewidth $\Delta f_{\rm gal} \sim 10^{-3}\,f$ set by the Galactic velocity dispersion. This characteristic makes the signal distinguishable from instrumental lines and amenable to targeted narrowband searches. While our analysis has focused on the smooth NFW halo, substructures or axion miniclusters could further enhance the effective flux, leading to stronger bounds or transient, ultra-narrow signals. Complementary observational probes, such as gravitational birefringence induced by the CS term, will provide independent tests of the same interaction, see Sec.~\ref{sec:introduction}.

\section{Discussion}
\label{sec:discussion}

While we have derived constraints under a set of well-motivated assumptions, the broader phenomenology and its connections to complementary experimental programs deserve further discussion. A key observational pathway is the search for parity-violating gravitational birefringence, which arises generically in CS gravity and is already being constrained in binary coalescence events~\cite{Yunes:2010yf, Alexander:2017jmt, Yamada:2020zvt, Okounkova:2021xjv}. A birefringent signature with time- or source-dependent structure would most naturally indicate a propagation effect rather than a Galactic DM signal, whereas the detection of a persistent, narrow spectral line would point instead to axion decay. Observing both signatures simultaneously would strongly suggest a common underlying parity-violating interaction. Our results can be re-expressed in the standard language of stochastic backgrounds through the spectral density
\begin{equation}
    \Omega_{\rm GW}h^2 \equiv \frac{1}{\rho_c}\frac{{\rm d}\rho_{\rm GW}}{{\rm d}\ln f}\,,
\end{equation}
where $\rho_c$ is the critical density. Mapping the line-like signal at frequency $f\simeq m_\phi/(4\pi)$ into this quantity provides a one-to-one relation between the axion lifetime and the induced GW density. In this representation, the O4 sensitivity corresponds to $\Omega_{\rm GW}h^2 \sim 10^{-9}$ at $f\sim 10^2$\,Hz, consistent with our Fig.~\ref{fig:bounds_interpolation}.

Our analysis assumes a smooth NFW halo, but the Galactic DM distribution may contain substructures, streams, or clumps, which can significantly enhance the $D$-factor in certain sky directions. Incorporating kinematic data from surveys such as Gaia will improve these estimates. A more accurate, and potentially larger, $D$-factor would imply a stronger predicted strain for a fixed coupling $\alpha$, and therefore, null results from interferometers could be used to place even tighter constraints. The existence of dense axion substructures such as miniclusters~\cite{Hogan:1988mp, Kolb:1993zz} or axion stars~\cite{Visinelli:2017ooc, Hertzberg:2018zte} presents a more complex but potentially powerful observational scenario. These objects possess two key features: an extremely low velocity dispersion, which would produce a spectrally narrow signal $\Delta f_{\rm sub} \ll  \Delta f_{\rm gal}$, and a very high local density. The detection prospects depend critically on their abundance, density, and spatial distribution. A nearby, dense minicluster could, in principle, produce a transient, spectrally ultra-narrow line that is significantly brighter in its specific sky location than the isotropic smooth-halo signal. The narrow linewidth would allow narrowband searches, potentially leading to very strong constraints if the object is found. However, this requires a challenging all-sky direction-dependent search. Furthermore, the random motion of these objects introduces a transient nature to the signal. Conversely, the non-detection of such bright, narrow features in high-resolution spectrograms could be used to constrain the abundance of the densest miniclusters~\cite{ADMX:2024pxg}.

In addition to the Galactic contribution, one expects a diffuse extragalactic background from axion decay in halos across cosmic history. The amplitude of this component can be estimated by integrating over the halo mass function and redshift distribution. For decays into monochromatic gravitons, the extragalactic signal is broadened by cosmological redshift and tends to produce a smooth stochastic background rather than a narrow line. While the overall intensity may be comparable in order of magnitude to the Galactic halo contribution, the loss of spectral sharpness substantially reduces its detectability in narrowband searches. Thus, for line-like signals, the Galactic halo dominates, although the extragalactic component remains of interest for stochastic searches.

Another important caveat to our analysis is that we have focused on perturbative axion decays, whereas non-perturbative processes such as parametric resonance can dominate the instability rate in some regions of parameter space. In particular, resonance effects can dramatically enhance the decay probability, leading to bounds on the CS coupling that are orders of magnitude stronger than those derived from the perturbative rate~\cite{Arza:2018dcy, Arza:2020eik}. In particular, Ref.~\cite{Jung:2020aem} shows that coherent axion fields can resonantly amplify chirping GWs from binaries as they traverse the halo, producing constraints far stronger than ours in the window $m_\phi \sim 10^{-13}\textrm{--}10^{-11}\,$eV. However, these limits rely on assumptions about coherence length, path length, and waveform modeling, while our spontaneous-decay constraints require only the presence of a Galactic axion background. In regions where resonance applies, the parameter space is more restricted than shown in Fig.~\ref{fig:bounds_interpolation}, whereas in regimes where coherence breaks down, our results remain conservative and robust. A systematic inclusion of these non-perturbative effects is presented in upcoming work.

Several distinctive signatures could help distinguish a true signal from instrumental artifacts. The expected linewidth of the signal is set by the Galactic velocity dispersion, naturally producing a width $\Delta f_{\rm gal}/f \sim 10^{-3}$. Annual modulation at the $\sim 5$–10\% level is expected from the Earth’s motion through the halo, providing a robust discriminant of a DM origin. Most strikingly, the circular polarization asymmetry predicted by the CS coupling would serve as a smoking gun if measurable. This is a prime target for third-generation ground-based detectors such as the ET and Cosmic Explorer, whose multi-arm configurations can resolve polarization. Finally, on the theoretical side, our photophobic axion assumption, while restrictive, has natural motivation in several beyond SM frameworks, including string-inspired models where the gravitational CS coupling is comparatively enhanced. Relaxing this assumption, and allowing EM channels to compete, will clarify how robust the bounds on $\alpha$ remain once realistic model-building constraints are imposed.

The mapping between axion mass and GW frequency implies that other classes of GW experiments probe distinct mass windows. For example, while our analysis has focused on ground-based interferometers, space-based detectors such as Laser Interferometer Space Antenna (LISA)~\cite{LISA:2017pwj}, DECi-hertz Interferometer Gravitational Wave Observatory (DECIGO)~\cite{Seto:2001qf}, and Big Bang Observer (BBO)~\cite{Harry:2006fi} will probe a complementary mass window. Their peak sensitivities in the mHz–Hz band correspond to axion masses in the range $10^{-17} \lesssim m_\phi/{\rm eV} \lesssim 10^{-14}$, well below the reach of terrestrial facilities. In this regime, the CS coupling could generate a persistent monochromatic signal analogous to that discussed here, potentially accompanied by birefringence and polarization asymmetries. Incorporating space-based data would therefore extend the accessible parameter space by several orders of magnitude in $m_\phi$, providing a powerful complement to ground-based searches. In addition, pulsar timing arrays (PTAs) are sensitive in the nHz band, corresponding to axion masses $m_\phi \sim 10^{-23}\textrm{--}10^{-20}$\,eV. In this window, the predicted extragalactic signal would be spectrally broad due to cosmological redshifting, but the Galactic halo contribution could in principle produce a quasi-monochromatic feature. Because the expected linewidth, $\Delta f_{\rm gal} / f \sim 10^{-3}$, is well below the frequency resolution set by the observation time, PTAs would register the signal as effectively a single Fourier bin. This places the signal conceptually closer to the continuous-wave sources already targeted in PTA analyses. A dedicated narrowband search would therefore be feasible in principle, though it would need to include correlated monochromatic systematics. Existing PTA bounds on stochastic backgrounds already reach $\Omega_{\rm GW} h^2 \sim 10^{-10}\textrm{--}10^{-11}$ in this band, suggesting that competitive constraints on the axion–CS coupling could be derived. More generally, any GW facility with sensitivity to stochastic or continuous signals can probe the axion–CS interaction in its corresponding frequency range: PTAs at the lowest masses, LISA/DECIGO in the intermediate window, and terrestrial interferometers at higher masses.

\section{Conclusions}
\label{sec:conclusions}

In this work, we investigated the decay of non-relativistic axion-like particles into gravitons via the dynamical CS interaction, focusing on the regime where this decay dominates. Modeling the Galactic axion distribution with a NFW profile, we computed the expected GW flux and compared it to the sensitivities of the publicly available data from the LIGO O4 run, as well as the projected sensitivity of ET. From this analysis we derived robust direct terrestrial forecasts on the axion–graviton coupling $\alpha$, surpassing the cosmological stability requirement for axion masses $m_\phi \lesssim 10^{-11}$\,eV.  In particular, the O4 sensitivity improves upon the cosmological stability limit by up to four orders of magnitude in the vicinity of $m_\phi \sim 10^{-12}$\,eV. These results establish ground-based interferometers as precision probes of axion DM with purely gravitational couplings, including \emph{photophobic} axion models.

Looking ahead, our results establish GW interferometers as precision probes of parity-violating gravity and physics beyond the SM. Sensitivity gains from third-generation detectors, together with complementary frequency coverage from space-based missions such as LISA, BBO or DECIGO, will push these bounds deeper and may reveal a cosmological contribution to the stochastic background from axion decay. Future work should relax the photophobic assumption to quantify the role of EM decay channels, and incorporate the impact of DM substructures or axion miniclusters, which could generate transient and spectrally sharp signals. Another promising direction is the study of axion condensates around compact objects, where the CS coupling may leave measurable imprints on quasinormal modes, ringdowns, or tidal dynamics. Collectively, these avenues outline a broader research program in which GW astronomy becomes a direct probe of axion physics and parity-violating interactions.

\vspace{.3cm}
\begin{acknowledgments}
We gratefully thank Ariel Arza and Edoardo Vitagliano for reading a preliminary version of the draft and for suggesting comments. We also thank Javi Serra for pointing out the discussion on EFT bounds. We acknowledge support by Istituto Nazionale di Fisica Nucleare (INFN) through the Commissione Scientifica Nazionale 4 (CSN4) Iniziativa Specifica ``Quantum Universe'' (QGSKY). LV also thanks the National Natural Science Foundation of China (NSFC) through the grant No.\ 12350610240 ``Astrophysical Axion Laboratories'', along with the Tsung-Dao Lee Institute and the Xplorer Symposia Organization Committee of the New Cornerstone Science Foundation for hospitality during the final stages of this work. This publication is based upon work from the COST Actions ``COSMIC WISPers'' (CA21106) and ``Addressing observational tensions in cosmology with systematics and fundamental physics (CosmoVerse)'' (CA21136), both supported by COST (European Cooperation in Science and Technology).
\end{acknowledgments}

\setlength{\bibsep}{4pt}
\bibliographystyle{apsrev4-1}
\bibliography{bibliography}

\end{document}